# Generalized Minimum Error with Fiducial Points Criterion for Robust Learning(This paper has been submitted in IEEE, it may be accepted later.)

Haiquan Zhao, *Senior Member, IEEE*, Yuan Gao, and Yingying Zhu

*Abstract*—The conventional Minimum Error Entropy criterion (MEE) has its limitations, showing reduced sensitivity to error mean values and uncertainty regarding error probability density function locations. To overcome this, a MEE with fiducial points criterion (MEEF), was presented. However, the efficacy of the MEEF is not consistent due to its reliance on a fixed Gaussian kernel. In this paper, a generalized minimum error with fiducial points criterion (GMEEF) is presented by adopting the Generalized Gaussian Density (GGD) function as kernel. The GGD extends the Gaussian distribution by introducing a shape parameter that provides more control over the tail behavior and peakedness. In addition, due to the high computational complexity of GMEEF criterion, the quantized idea is introduced to notably lower the computational load of the GMEEF-type algorithm. Finally, the proposed criterions are introduced to the domains of adaptive filter, kernel recursive algorithm, and multilayer perceptron. Several numerical simulations, which contain system identification, acoustic echo cancellation, times series prediction, and supervised classification, indicate that the novel algorithms' performance performs excellently.

*Index Terms*—Generalized gaussian density, generalized minimum error with fiducial points criterion, quantization, kernel recursive algorithm, acoustic echo cancellation, multilayer perceptron.

## I. INTRODUCTION

ENTROPY, a latest local similarity measure in information theoretic learning (ITL), is utilized in adaptive filters (AF) [1] as a means to describe the comprehensive information content of the signal. By mapping the second-order statistics to the feature space, entropy proves to be more efficient in handling impulsive noise compared to merely replacing the error moments [1]. However, traditional maximum correntropy criterion (MCC) performs poorly in response to nonlinear and non-Gaussian signal interference. Countering this situation, MEE criterion [3][4] has attracted a lot of attention as a significant cost function in ITL. It captures higher-order statistics of the input data and is robust to outliers [5]. A variety of properties of MEE have been reported in [6] to well-established it. This has led the MEE to demonstrate extensive applicability in a range of areas, such as clustering problem [7], and machine learning [8], classification problems [9], face recognition [10] as well.

Typical MEE-based AF algorithms have widely applied in linear [11][12] and nonlinear [13][14] systems. However, the MEE criterion still has some drawbacks in that it cannot accurately localize the error probability density function (PDF). Due to the fact that MEE is shift-invariant, a bias must be added to achieve zero-mean error for the training data set. The MCC in ITL has a probabilistic meaning, i.e., it maximizes the error PD at the origin [15][16]. Therefore, a weighted combination of MEE and MCC has been presented and named MEEF, which is superior to MEE and MCC and does not require adding bias after training [17].

Although the MEEF with default Gaussian kernel has compelling approximation ability, it is not impeccable [18]. The Gaussian kernel is both smooth and strictly positive-definite, while also representing maximum ambiguity or lack of information regarding all other aspects [19][20]. To enhance the MEEF, the paper introduces the GMEEF, which employs the GGD function as its kernel [21]. The GGD has found applications in various fields, including image processing, signal processing, and statistics. It is particularly useful when modeling data with varying characteristics, where the Gaussian distribution might not adequately capture the underlying distribution. The proposed GMEEF incorporates higher-order absolute moments, including the second-order, of the error signal in the feature space, which enhances its ability to handle multiple intricate noise environments. Additionally, to reduce the computational load caused by the GMEE component in the GMEEF criterion, a new criterion called QGMEEF is introduced, which quantizes the error set using a specific quantization scheme [8].

Finally, some expansions of the proposed new criterions are proposed. Firstly, kernel recursive GMEEF (KRGMEEF) are developed to test the validity of the presented criterion in nonlinear and non-stationary systems [14]. Secondly, an acoustic echo cancellation (AEC) based on GMEEF criterion is presented to verify the performance of the criterion in echo cancellation field [22]. Finally, a multilayer perceptron (MLP) network-based [23] the GMEEF criterion is trained to be used in classification. The main achievements are concluded in this paper as below.

1) Inspired by MEEF criterion and GGD function, a weighted combination of GMEE and GMCC, namely GMEEF

This work was partially supported by National Natural Science Foundation of China (grant: 62171388, 61871461, 61571374), and Fundamental Research Funds for the Central Universities (grant: 2682021ZTPY091).

The authors are with the Key Laboratory of Magnetic Suspension Technology and Maglev Vehicle, Ministry of Education, Southwest Jiaotong University, Chengdu 610031, China, and with the School of Electrical Engineering, Southwest Jiaotong University, Chengdu, 610031, China (e-mail: hqzhao_swjtu@126.com, gaoyuan1595_2@126.com, zhuyingying_629@126.com). (Corresponding author: Haiquan Zhao.)

is proposed. The GMEEF criterion combines the benefits of both GMCC and GMEE criteria. It captures higher-order statistics of error and maintain sensitivity to error mean in biased or non-Gaussian signals. Some simply properties are discussed.

2) The concept of quantization is introduced, giving rise to the proposition of the QGMEEF criterion. Compared with the GMEEF criterion, the QGMEEF criterion can greatly reduce the computational complexity and has the virtually unchanged performance.

3) The new AF, kernel recursive least squares (KRLS), and MLP models-based these criteria are implemented. To evaluate the effectiveness of these new criteria, various simulations are performed, including system identification, AEC, time series prediction, and supervised classification. The results of these simulations provide evidence of the performance and reliability of these new criteria.

The content of this article is summarized as below: the GMEEF and QGMEEF criteria and some analysis are presented in Section 2. In Section 3, the new AF, KRLS algorithms, and MLP network are illustrated. Simulations are performed in Section 4. Section 5 concludes all the paper.

## II. PROPOSED CRITERIONS

Within this section, we incorporate the GGD function into the ITL, introducing the GMEEF along with its quantized counterpart.

*A. The GMEEF Criterion*

In ITL, $\mathbf{e} = \mathbf{Y} - \mathbf{X}$, which is assumed as the error variable, is represented using 2-order Renyi's entropy [4].

$$H_2(\mathbf{e}) = \log V_2(\mathbf{e}) \tag{1}$$

where $V_2(\mathbf{e})$, namely the quadratic information potential (QIP), is indicated as

$$V_2(\mathbf{e}) = \int p_\mathbf{e}^2(x)dx = E[p(\mathbf{e})] \tag{2}$$

where $p(\mathbf{e})$ and $p_\mathbf{e}(x)$ indicate the PDF of $\mathbf{e}$ and its value at the point $x$, respectively. Due to the exact distribution of the error variable in (2), $p_\mathbf{e}(x)$ is not known in practice. To estimate the distribution, the Parzen's window estimator can be employed [4]. $p_\mathbf{e}(x)$ is obtained by applying.

$$\hat{p}_\mathbf{e}(x) = \frac{1}{L}\sum_{i=1}^{L} G_\sigma(x - e_i) \tag{3}$$

where $G_\sigma(x) = \exp(-x^2/2\sigma^2)$ defines the Gaussian kernel with size $\sigma$ [4]. $\mathbf{e} = \{e_i\}_{i=1}^{L}$ expresses the error set with $L$ error samples. An estimate of $V_2^{MEE}(\mathbf{e})$ can be calculated by

$$\widehat{V}_2^{MEE}(\mathbf{e}) = \frac{1}{L}\sum_{i=1}^{L}\hat{p}(e_i) = \frac{1}{L^2}\sum_{i=1}^{L}\sum_{j=1}^{L}G_\sigma(e_i - e_j) \tag{4}$$

The MEE criterion can be contained by maximizing the QIP. However, the entropy remains unchanged with respect to the distribution's mean, potentially resulting in a non-zero mean error, even as the MEE-type algorithm converges towards a certain value. In general, to obtain zero-mean error, it is also necessary to manually add bias to the model to accomplish the filtering purpose of minimizing the error to zero. The MCC criterion can be interpreted probabilistically as maximizing the density of error probabilities at the origin.

Typically, the conventional error entropy and correntropy adopt a Gaussian kernel function, which limits their ability to freely change the shape of the distribution as they can only handle Gaussian noise. To enhance the conventional error entropy, we introduce the GGD function into the ITL framework. This integration results in a novel criterion that combines the benefits of both error entropy and correntropy.

The GGD function is defined as [20]:

$$G_{\alpha,\beta}(e) = \frac{\alpha}{2\beta\Gamma(1/\alpha)}\exp(-\left|\frac{e}{\beta}\right|^\alpha) \tag{5}$$

where $\alpha$, $\beta$ describe the exponential decay rate and the dispersion of the distribution, individually. $\Gamma(\bullet)$ means the gamma function $\Gamma(\bullet) = \int_0^\infty x^{c-1}e^{-x}dx, c > 0$. The generalized function discussed in [25] is closely connected to the generalized gamma distribution. When the shape parameter $\alpha$ is assigned a value of one or two, the GGD distribution transforms into the Laplacian or Gaussian distribution, respectively.

Then, adopting the GGD as kernel, the QIP of GMEEF is defined as [17]:

$$\begin{aligned}\widehat{V}_{\alpha,\beta}^{GMEEF}(\mathbf{e}) &= \widehat{V}_{\alpha_1,\beta_1}^{GMCC}(\mathbf{e}) + \widehat{V}_{\alpha_2,\beta_2}^{GMEE}(\mathbf{e}) \\ &= \frac{\lambda}{L}\sum_{i=1}^{L}G_{\alpha_1,\beta_1}(e_i) + \frac{1-\lambda}{L^2}\sum_{i=1}^{L}\sum_{j=1}^{L}G_{\alpha_2,\beta_2}(e_i - e_j)\end{aligned} \tag{6}$$

Then, some properties are discussed.

**Property 1.** $\widehat{V}_2^{GMEEF}(\mathbf{e})$ is symmetric, it can be seen from the symmetry of $\widehat{V}_2^{GMCC}(\mathbf{e})$ and $\widehat{V}_2^{GMEE}(\mathbf{e})$ [26].

$$\begin{aligned}\widehat{V}_{\alpha,\beta}^{GMEEF}(-\mathbf{e}) &= \widehat{V}_{\alpha_1,\beta_1}^{GMCC}(-\mathbf{e}) + \widehat{V}_{\alpha_2,\beta_2}^{GMEE}(-\mathbf{e}) \\ &= \frac{\lambda}{L}\sum_{i=1}^{L}G_{\alpha_1,\beta_1}(-e_i) + \frac{1-\lambda}{L^2}\sum_{i=1}^{L}\sum_{j=1}^{L}G_{\alpha_2,\beta_2}(-(e_i - e_j)) \\ &= \frac{\lambda}{L}\sum_{i=1}^{L}G_{\alpha_1,\beta_1}(e_i) + \frac{1-\lambda}{L^2}\sum_{i=1}^{L}\sum_{j=1}^{L}G_{\alpha_2,\beta_2}(e_i - e_j) \\ &= \widehat{V}_{\alpha_1,\beta_1}^{GMCC}(\mathbf{e}) + \widehat{V}_{\alpha_2,\beta_2}^{GMEE}(\mathbf{e}) \\ &= \widehat{V}_{\alpha,\beta}^{GMEEF}(\mathbf{e})\end{aligned} \tag{7}$$

**Property 2.** $\widehat{V}_{\alpha,\beta}^{GMEEF}(\mathbf{e})$ is both positive and bounded, satisfies $0 < \widehat{V}_{\alpha,\beta}^{GMEEF}(\mathbf{e}) \leq \lambda\alpha_1/2\beta_1\Gamma(1/\alpha_1) + (1-\lambda)\alpha_2/2\beta_2\Gamma(1/\alpha_2)$, and $\widehat{V}_{\alpha,\beta}^{GMEEF}(\mathbf{e}) = \lambda\alpha_1/2\beta_1\Gamma(1/\alpha_1) + (1-\lambda)\alpha_2/2\beta_2\Gamma(1/\alpha_2)$ if and only if $\mathbf{e} = 0$.

In detail, $\widehat{V}_{\alpha_1,\beta_1}^{GMCC}(\mathbf{e})$ and $\widehat{V}_{\alpha_2,\beta_2}^{GMEE}(\mathbf{e})$ are both positive and bounded. The GGD function still satisfies some properties of the Gaussian kernel function, i.e.

$$0 < \widehat{V}_{\alpha_1,\beta_1}^{GMCC}(\mathbf{e}) \leq \frac{\alpha_1}{2\beta_1\Gamma(1/\alpha_1)} \tag{8}$$

$$0 < \widehat{V}_{\alpha_2,\beta_2}^{GMEE}(\mathbf{e}) \leq \frac{\alpha_2}{2\beta_2\Gamma(1/\alpha_2)} \tag{9}$$

Thus $\hat{V}_{\alpha,\beta}^{GMEEF}(\mathbf{e})$, which is structured as a weighted average of $\hat{V}_{\alpha_1,\beta_1}^{GMCC}(\mathbf{e})$, $\hat{V}_{\alpha_2,\beta_2}^{GMEE}(\mathbf{e})$ is bounded, and also under this condition, the range of $\hat{V}_{\alpha,\beta}^{GMEEF}(\mathbf{e})$ should conform to the weighted average of $\hat{V}_{\alpha_1,\beta_1}^{GMCC}(\mathbf{e})$, $\hat{V}_{\alpha_2,\beta_2}^{GMEE}(\mathbf{e})$. So, property 2 holds.

**Property 3**. $\hat{V}_2^{GMEEF}(\mathbf{e})$ can be further denoted as (10) via the Taylor expansion.

$$\hat{V}_2^{GMEEF}(\mathbf{e}) = \frac{\lambda \alpha_1}{2\beta_1 \Gamma(1/\alpha_1)} \sum_{k=0}^{\infty} \frac{(-1)^k}{k! \beta_1^{\alpha_1 k}} E[|\mathbf{e}|^{\alpha_2 k}]$$
$$+ \frac{(1-\lambda)\alpha_2}{2L\beta_2 \Gamma(1/\alpha_2)} \sum_{i=1}^{L} \sum_{k=0}^{\infty} \frac{(-1)^k}{k! \beta_2^{\alpha_2 k}} E[|e_i - \mathbf{e}|^{\alpha_2 k}] \quad (10)$$

*B. The QMEEF Criterion*

According to (6), the QIP of the GMEE part can be computed using the double summation method. However, this method can be computationally expensive, particularly for large datasets, leading to a significant computational burden in obtaining the QIP. Thus, referring to previous studies [8], quantized GMEEF is proposed to quantize the GMEE part.

$$\hat{V}_{\alpha_2,\beta_2}^{GMEE}(\mathbf{e}) \approx \hat{V}_{\alpha_2,\beta_2}^{QGMEE}(\mathbf{e})$$
$$= \frac{1-\lambda}{L^2} \sum_{i=1}^{L} \sum_{j=1}^{L} G_{\alpha_2,\beta_2}(e_i - Q[e_j, \gamma]) \quad (11)$$
$$= \frac{1-\lambda}{L^2} \sum_{i=1}^{L} \sum_{h=1}^{H} H_h G_{\alpha_2,\beta_2}(e_i - o_h)$$

Getting a codebook $\mathbf{O} = \{o_1, o_2, \ldots, o_H \in \mathbb{R}^1\}$ ($H \leq L$) set with quantization operator $Q[e_i, \gamma] \in K$ [8] ($\gamma$ denotes the quantization threshold). $H$ expresses the number of codes in error set. Therefore, the cost functions of GMEEF and QGMEEF are shows as follows:

$$\begin{cases} J_{GMEEF} = \arg\max\{\frac{\lambda}{L}\sum_{i=n}^{n+L-1}G_{\alpha_1,\beta_1}(e_i) + \frac{1-\lambda}{L^2}\sum_{i=n}^{n+L-1}\sum_{j=n}^{n+L-1}G_{\alpha_2,\beta_2}(e_i - e_j)\} \\ J_{QGMEEF} = \arg\max\{\frac{\lambda}{L}\sum_{i=n}^{n+L-1}G_{\alpha_1,\beta_1}(e_i) + \frac{1-\lambda}{L^2}\sum_{i=n}^{n+L-1}\sum_{h=1}^{H}H_h G_{\alpha_2,\beta_2}(e_i - o_h)\} \end{cases} \quad (12)$$

When $H = L$, the QGMEEF criterion reduces to the ordinary GMEEF criterion.

One of the main challenges lies in designing the vector quantizer (VQ). In this study, an online VQ method is employed, where the codebook is trained directly from online samples [27]. This online VQ method has the capability to adaptively grow. The details of the online VQ method in $\mathbb{R}$ are provided in Table I. In this method, $\mathbf{O}_j(i-1)$ represents the $j$ th code vector of the codebook $\mathbf{O}(i-1)$, and $\|\cdot\|$ expresses the Euclidean norm in $\mathbb{R}$.

TABLE I
ONLINE VQ METHOD

| |
|---|
| **Input:** $\{\mathbf{a}(i) \in \mathbb{R}^{N \times 1}\}, i = 1, \ldots, N$ |
| **Initialization:** $\varepsilon \geq 0$, and $\mathbf{O}(1) = \{\mathbf{a}(1)\}$ |
| **While** $\{\mathbf{a}(i)\}(i > 1)$ **available do** |
|   1) Compare distance: $dis(\mathbf{a}(i), \mathbf{O}(i-1)) = \min_{1 \leq j \leq size(\mathbf{O}(i-1))} \|\mathbf{a}(i) - \mathbf{a}_j(i-1)\|$ |
|   2) If $dis(\mathbf{a}(i), \mathbf{O}(i-1)) \leq \varepsilon$, $\mathbf{O}(i) = \mathbf{O}(i-1)$, and $\mathbf{a}_q(i) = \mathbf{O}_{j^*}(i-1)$, where $j^* = \arg\min_{1 \leq j \leq size(\mathbf{O}(i-1))} \|\mathbf{O}(i) - \mathbf{O}_j(i-1)\|$ |
|   3) Else, $\mathbf{O}(i) = \{\mathbf{O}(i-1), \mathbf{a}(i)\}$, and $\mathbf{a}_q(i) = \mathbf{a}(i)$ |
| **End** |

## III. LEARNING ALGORITHM

In this section, the AF algorithm based on GMEEF is proposed in Part A. In Part B, the AF algorithm based on QGMEEF is introduced. The KRGMEEF algorithm is concluded in Part C. And the MLP network based on GMEEF (MLP-GMEEF) is given in Part D.

*A. AF algorithm-based GMEEF*

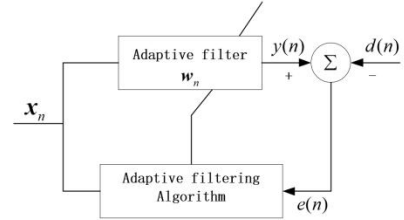

**Fig. 1.** AF algorithm schematic.

Fig.1 shows a schematic of the AF algorithm. Assuming that a linear regression model (LRM) $d(n) = \mathbf{w}_n^T \mathbf{x}_n + v(n)$ has been identified, where $\mathbf{w}_n = [w(n), w(n-1), \ldots, w(n-L+1)]^T$ and $\mathbf{x}_n = [x(n), x(n-1), \ldots, x(n-L+1)]^T$ denote the unknown weight vector and the input signal, individually. The error signal $e(n) = d(n) - \mathbf{w}_n^T \mathbf{x}_n$ is indicated.

From (12), the iteration formula that represents the weight vector of the GMEEF algorithm is

$$\mathbf{w}_{n+1} = \mathbf{w}_n + \mu \nabla J_{GMEEF}(\mathbf{w}_n)$$
$$= \mathbf{w}_n + \mu\{\frac{\lambda}{L}\sum_{i=n}^{n+L-1} G_{\alpha_1,\beta_1}(e(i))|e(i)|^{\alpha_1-1} sign(e(i))\mathbf{x}_i$$
$$+ \frac{1-\lambda}{L^2}\sum_{i=n}^{n+L-1}\sum_{j=n}^{n+L-1}\begin{bmatrix} G_{\alpha_2,\beta_2}(\triangle e_{i,j})|\triangle e_{i,j}|^{\alpha_2-1} \\ sign(\triangle e_{i,j})(\triangle \mathbf{x}_{i,j}) \end{bmatrix}\} \quad (13)$$
$$= \mathbf{w}_n + \mu \mathbf{X}_n\{\frac{\lambda}{L}\mathbf{R}_n^T + \frac{1-\lambda}{L^2}(\mathbf{P}_n^T - \mathbf{Q}_n^T)\}$$

$$\begin{cases} \boldsymbol{X}_n = [\boldsymbol{x}_i \quad \boldsymbol{x}_{n+1} \quad \cdots \quad \boldsymbol{x}_{n+L-1}] \\ \boldsymbol{R}_n = [r_{n:n} \quad r_{n:n+1} \quad \cdots \quad r_{n:n+L-1}] \\ \boldsymbol{P}_n = [p_{n:n} \quad p_{n:n+1} \quad \cdots \quad p_{n:n+L-1}] \\ \boldsymbol{Q}_n = [q_{n:n} \quad q_{n:n+1} \quad \cdots \quad q_{n:n+L-1}] \\ r_{n:i} = G_{\alpha_1,\beta_1}(e(i))|e(i)|^{\alpha_1-1}sign(e(i)) \\ p_{n:i} = \sum_{j=n}^{n-L+1}[G_{\alpha_2,\beta_2}(\Delta e_{i,j})|\Delta e_{i,j}|^{\alpha_2-1}sign(\Delta e_{i,j})] \\ q_{n:i} = \sum_{j=n}^{n-L+1}[G_{\alpha_2,\beta_2}(\Delta e_{i,j})|\Delta e_{i,j}|^{\alpha_2-1}sign(\Delta e_{j,i})] \end{cases} \quad (14)$$

where $\Delta a_{i,j} = a(i) - a(j)$, $sign(\cdot)$ denotes the symbolic function and $\mu > 0$ indicates the step size of GMEEF algorithm. Obviously, when $\alpha_1 = \alpha_2 = 2$, the GMEEF criterion becomes the original MEEF criterion [17].

Conducting a convergence analysis of GMEEF is essential. To streamline the analytical treatments, need to take into account certain common assumptions [20][21].

*Assumption 1:* The priori errors $\varepsilon_{a:n}$ and posteriori errors $\varepsilon_{p:n}$ are assumed to be independence of the noise $v(n)$.

*Assumption 2:* $x(i)$ and $v(j)$ are uncorrelated, when $i \neq j$. The variance of input signal is $\sigma_x^2$.

From (13) and (14), (15) can be got.

$$\boldsymbol{\omega}_{n+1} = \boldsymbol{\omega}_n - \mu \boldsymbol{X}_n \{\frac{\lambda}{L}\boldsymbol{R}_n^T + \frac{1-\lambda}{L^2}(\boldsymbol{P}_n^T - \boldsymbol{Q}_n^T)\} \quad (15)$$

where $\boldsymbol{\omega}_n = \boldsymbol{w}_0 - \boldsymbol{w}_n$, $\boldsymbol{w}_0$ denotes the optimum tap weight vector, $\varepsilon_{a:n}$ and $\varepsilon_{p:n}$ of the GMEEF algorithm are expressed as

$$\begin{cases} \varepsilon_{a:n} = \boldsymbol{X}_n^T \boldsymbol{\omega}_n \\ \varepsilon_{p:n} = \boldsymbol{X}_n^T \boldsymbol{\omega}_{n+1} \end{cases} \quad (16)$$

where $\varepsilon_{a:n} = [e_a(n) \quad e_a(n+1) \quad \cdots \quad e_a(n+L-1)]$. For simplicity, a number of expressions can be used.

$$\begin{cases} \boldsymbol{D}_n = [d(n) \quad d(n+1) \quad \cdots \quad d(n+L-1)]^T \\ \quad = [\boldsymbol{D}_{n-1}(2:L) \quad d(n+L-1)]^T \\ \quad = [\boldsymbol{D}_{n-1}(2:L) \quad d_L]^T \\ \boldsymbol{V}_n = [v(n) \quad v(n+1) \quad \cdots \quad v(n+L-1)]^T = [\boldsymbol{V}_{n-1}(2:L) \quad v_L]^T \\ \boldsymbol{\varepsilon}_n = [e(n) \quad e(n+1) \quad \cdots \quad e(n+L-1)]^T \\ \quad = [\boldsymbol{\varepsilon}_{n-1}(2:L) \quad e_L]^T \\ \quad = \boldsymbol{D}_n - \boldsymbol{X}_n^T \boldsymbol{w}_n \end{cases} \quad (17)$$

Combining (16) and the third term of (17), $\varepsilon_n = \varepsilon_{a:n} + \boldsymbol{V}_n$ can be obtained. Left multiply both sides of (15) by $\boldsymbol{X}_n^T$, $\varepsilon_{p:n} = \varepsilon_{a:n} - \mu \boldsymbol{X}_n^T \boldsymbol{X}_n \{\frac{\lambda}{L}\boldsymbol{R}_n^T + \frac{1-\lambda}{L^2}(\boldsymbol{P}_n^T - \boldsymbol{Q}_n^T)\}$. So (18) can be obtained with the assumption that matrix $\boldsymbol{X}_n^T \boldsymbol{X}_n$ is invertible.

$$(\boldsymbol{X}_n^T \boldsymbol{X}_n)^{-1}(\varepsilon_{a:n} - \varepsilon_{p:n}) = \mu\{\frac{\lambda}{L}\boldsymbol{R}_n^T + \frac{1-\lambda}{L^2}(\boldsymbol{P}_n^T - \boldsymbol{Q}_n^T)\} \quad (18)$$

Put (18) into (15), and $\boldsymbol{\omega}_{n+1} = \boldsymbol{\omega}_n - \boldsymbol{X}_n(\boldsymbol{X}_n^T\boldsymbol{X}_n)^{-1}(\varepsilon_{a:n} - \varepsilon_{p:n})$. Then, square both sides and take the expectation both sides to obtain

$$E[\|\boldsymbol{\omega}_{n+1}\|^2] = E[\|\boldsymbol{\omega}_n\|^2] - 2E[\varepsilon_{a:n}^T(\boldsymbol{X}_n^T\boldsymbol{X}_n)^{-1}(\varepsilon_{a:n} - \varepsilon_{p:n})] \\ + E\begin{bmatrix} [(\boldsymbol{X}_n^T\boldsymbol{X}_n)^{-1}(\varepsilon_{a:n} - \varepsilon_{p:n})]^T \times \\ \boldsymbol{X}_n^T\boldsymbol{X}_n(\boldsymbol{X}_n^T\boldsymbol{X}_n)^{-1}(\varepsilon_{a:n} - \varepsilon_{p:n})] \end{bmatrix} \quad (19)$$

Combining (19) with (18), (20) is calculated as

$$E[\|\boldsymbol{\omega}_{n+1}\|^2] = E[\|\boldsymbol{\omega}_n\|^2] - 2\mu E[\frac{\lambda}{L}\varepsilon_{a:n}^T\boldsymbol{R}_n^T + \frac{1-\lambda}{L^2}\varepsilon_{a:n}^T(\boldsymbol{P}_n^T - \boldsymbol{Q}_n^T)] \\ + \mu^2 E\begin{bmatrix} [\frac{\lambda}{L}\boldsymbol{R}_n + \frac{1-\lambda}{L^2}(\boldsymbol{P}_n - \boldsymbol{Q}_n)] \times \\ \boldsymbol{X}_n^T\boldsymbol{X}_n[\frac{\lambda}{L}\boldsymbol{R}_n^T + \frac{1-\lambda}{L^2}(\boldsymbol{P}_n^T - \boldsymbol{Q}_n^T)] \end{bmatrix} \quad (20) \\ = E[\|\boldsymbol{\omega}_n\|^2] - 2\mu E[\varepsilon_{a:n}^T\boldsymbol{\Omega}_n] + \mu^2 E[\boldsymbol{\Omega}_n\boldsymbol{X}_n^T\boldsymbol{X}_n\boldsymbol{\Omega}_n^T]$$

When the system reaches steady-state $E[\|\boldsymbol{\omega}_{n+1}\|^2] = E[\|\boldsymbol{\omega}_n\|^2]$, the stable range of step size can be obtained:

$$\mu \leq \frac{E[\varepsilon_{a:n}^T\boldsymbol{\Omega}_n]}{E[\boldsymbol{\Omega}_n^T]\boldsymbol{X}_n^T\boldsymbol{X}_n E[\boldsymbol{\Omega}_n]} \quad (21)$$

$$\boldsymbol{\Omega}_n = \frac{\lambda}{L}\boldsymbol{R}_n^T + \frac{1-\lambda}{L^2}(\boldsymbol{P}_n^T - \boldsymbol{Q}_n^T) \quad (22)$$

According to assumption 2, (21) can be rewritten as

$$\mu \leq \frac{E[\varepsilon_{a:n}^T\boldsymbol{\Omega}_n]}{E[\boldsymbol{\Omega}_n^T]E[\boldsymbol{X}_n^T\boldsymbol{X}_n]E[\boldsymbol{\Omega}_n]} \quad (23)$$

According to assumption 2, $E[\boldsymbol{X}_n^T\boldsymbol{X}_n] = M\sigma_x^2 I_L$ can be obtained. Therefore, (24) can be obtained.

$$\mu \leq \frac{E[\varepsilon_{a:n}^T\boldsymbol{\Omega}_n]}{M\sigma_x^2 E[\boldsymbol{\Omega}_n^T]E[\boldsymbol{\Omega}_n]} \quad (24)$$

So, $E[\|\boldsymbol{\omega}_n\|^2]$ is convergent since it satisfies (24).

$$0 < \mu \leq \frac{E[\varepsilon_{a:n}^T\boldsymbol{\Omega}_n]}{M\sigma_x^2 E[\boldsymbol{\Omega}_n^T]E[\boldsymbol{\Omega}_n]} \quad (25)$$

When $n \to \infty$, $e(n) \to v(n)$. $\boldsymbol{R}_n$, $\boldsymbol{P}_n$, and $\boldsymbol{Q}_n$ are represented as

$$\begin{cases} \hat{\boldsymbol{R}}_n = [\hat{r}_{n:n} \quad \hat{r}_{n:n+1} \quad \cdots \quad \hat{r}_{n:n+L-1}] \\ \hat{\boldsymbol{P}}_n = [\hat{p}_{n:n} \quad \hat{p}_{n:n+1} \quad \cdots \quad \hat{p}_{n:n+L-1}] \\ \hat{\boldsymbol{Q}}_n = [\hat{q}_{n:n} \quad \hat{q}_{n:n+1} \quad \cdots \quad \hat{q}_{n:n+L-1}] \\ \hat{r}_{n:i} = G_{\alpha_1,\beta_1}(v(i))|v(i)|^{\alpha_1-1}sign(v(i)) \\ \hat{p}_{n:i} = E\left[\sum_{j=n}^{n-L+1}[G_{\alpha_2,\beta_2}(\Delta v_{i,j})|\Delta v_{i,j}|^{\alpha_2-1}sign(\Delta v_{i,j})]\right] \\ \hat{q}_{n:i} = E\left[\sum_{j=n}^{n-L+1}[G_{\alpha_2,\beta_2}(\Delta v_{i,j})|\Delta v_{i,j}|^{\alpha_2-1}sign(\Delta v_{j,i})]\right] \end{cases} \quad (26)$$

When the system is ergodic in a general sense, the last two items of (26) is

$$\begin{cases} \hat{p}_{n:i} \approx \frac{1}{i}\sum_{z=1}^{i}\sum_{j=n}^{n-L+1}\left[G_{\alpha_2,\beta_2}(\Delta v_{z,j})|\Delta v_{z,j}|^{\alpha_2-1}sign(\Delta v_{z,j})\right] \\ \hat{q}_{n:i} \approx \frac{1}{i}\sum_{z=1}^{i}\sum_{j=n}^{n-L+1}\left[G_{\alpha_2,\beta_2}(\Delta v_{z,j})|\Delta v_{z,j}|^{\alpha_2-1}sign(\Delta v_{j,z})\right] \end{cases} \quad (27)$$

If the step-size satisfies equation (25), it is evident that the sequence of $E[\|\omega_n\|^2]$ will converge.

*B. AF algorithm-based QGMEEF*

For reducing the computational burden of the GMEEF-type algorithm, an emerging AF algorithm is introduced, which utilizes the QGMEEF learning criterion. In the LRM, the cost function is represented by the second term in equation (12). Consequently, the iteration formula for the weight vector in QGMEEF algorithm is denoted.

$$\begin{aligned} \mathbf{w}_{n+1} &= \mathbf{w}_n + \mu_q \nabla J_{QMEEF}(\mathbf{w}_n) \\ &= \mathbf{w}_n + \mu_q \{ \frac{\lambda}{L} \sum_{i=n}^{n-L+1} G_{\alpha_1,\beta_1}(e(i))|e(i)|^{\alpha_1-1} sign(e(i))\mathbf{x}_i \\ &\quad + \frac{1-\lambda}{L^2} \sum_{i=n}^{n-L+1} \sum_{h=1}^{H} \begin{bmatrix} G_{\alpha_2,\beta_2}(e(i)-c_h)|e(i)-o_h|^{\alpha_2-1} \\ sign(e(i)-e(j))H_h \mathbf{x}_i \end{bmatrix} \} \\ &= \mathbf{w}_n + \mu_q \mathbf{X}_n \{ \frac{\lambda}{L}\mathbf{P}_n^T + \frac{1-\lambda}{L^2}\mathbf{\Lambda}_n^T \} \\ &= \mathbf{w}_n + \mu_q \mathbf{R}_n \end{aligned} \quad (28)$$

$$\mathbf{R}_n = \frac{\lambda}{L}\mathbf{P}_n^T + \frac{1-\lambda}{L^2}\mathbf{\Lambda}_n^T \quad (29)$$

with

$$\begin{cases} \mathbf{\Lambda}_n = [\upsilon_n \quad \upsilon_{n+1} \quad \cdots \quad \upsilon_{n+L-1}] \\ \upsilon_i = \sum_{h=1}^{H}[H_h G_{\alpha_2,\beta_2}(e(i)-o_h)|e(i)-c_h|^{\alpha_2-1} sign(e(i)-o_h)] \end{cases} \quad (30)$$

where $\mu_q > 0$ indicates the step size of QMEEF algorithm and $\mathbf{x}_i = \mathbf{a}(i)$ in the online VQ method.

Similar as GMEEF, the stable range of step-size in QGMEEF can be obtained.

$$0 < \mu_q \leq \frac{E[\boldsymbol{\varepsilon}_{a:n}^T \mathbf{R}_n]}{M\sigma_x^2 E[\mathbf{R}_n^T]E[\mathbf{R}_n]} \quad (31)$$

*C. KRLS-Based GMEEF*

The kernel method is a powerful tool for nonparametric modeling that involves mapping input data from a lower-dimensional space $\mathbb{X}$ to a higher-dimensional space $\mathbb{F}$ utilizing a specific nonlinear transformation.

$$\varphi : \mathbb{X} \to \mathbb{F} \quad (32)$$

where $\varphi$ represents the transformed data in the kernel method. In linear AF, the kernel method is employed by introducing a kernel function $\kappa$. This kernel function enables the inner product operations in the linear adaptive filters to be replaced by the computation of $\kappa$ in the feature space, without requiring explicit knowledge of the underlying nonlinear mapping.

According to the Mercer theorem [28], for any given kernel function, a corresponding mapping $\kappa(\mathbf{x}_i, \mathbf{x}_j) = \varphi_i^T \varphi_j = \exp(-(\Delta \mathbf{x}_{i,j})^2 / 2\sigma^2)$, for $\forall \mathbf{x}_i, \mathbf{x}_j \in \mathbb{X}$ is utilized.

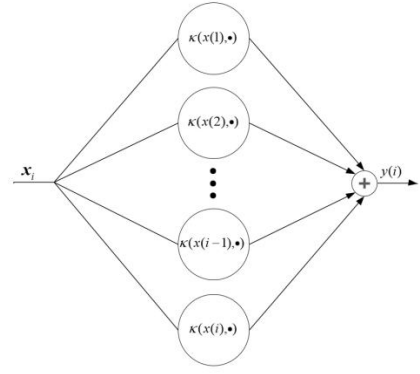

**Fig. 2.** Network topology of KRLS at iteration $i$.

In this part, an emerging KRLS algorithm-based the GMEEF criterion is presented. Fig. 2 illustrates the network topology of the KRLS algorithm. The kernel AF algorithm comprises two components: kernel least-mean-square (K-LMS) and KRLS algorithms. Although the KRLS offers superior performance compared to the KLMS, it comes at the expense of higher computational complexity. Examples of KRLS algorithms that utilize the kernel recursive GMEE (KRGMEE) and kernel recursive GMCC (KRGMCC) criteria have been previously developed [26][29]. Let us assume that equation (11) represents a nonlinear mapping. The cost function for the KRLS algorithm-based the GMEEF criterion is defined as below:

$$J_{KGMEEF} = \frac{\lambda}{L}\sum_{i=n}^{n-L+1}G_{\alpha_1,\beta_1}(e(i)) + \frac{1-\lambda}{L^2}\sum_{i=n}^{n-L+1}\sum_{j=n}^{n-L+1}G_{\alpha_2,\beta_2}(\Delta e_{i,j}) \\ -\frac{1}{2}\zeta_1\|\mathbf{w}_n\|^2 \quad (33)$$

where $e(n) = d(n) - f(\mathbf{x}_n)$, $\zeta_1$ denotes regularization factor, and the derivative $\partial J_{KGMEEF}/\partial \mathbf{w}_n$ is calculated.

$$\frac{\partial J_{KGMEEF}}{\partial \mathbf{w}_n} = \mathbf{\Phi}_L \mathbf{B}_L \mathbf{\Theta}_L - \zeta_1 \|\mathbf{w}_n\| \quad (34)$$

Setting $\partial J_{KGMEEF}/\partial \mathbf{w}_n = 0$ gives the solution

$$\mathbf{w} = (\mathbf{\Phi}_L \mathbf{B}_L \mathbf{\Phi}_L^T + \zeta_1 I)^{-1}\mathbf{\Phi}_L \mathbf{B}_L \mathbf{D}_L \quad (35)$$

(35) can be rewritten as

$$\mathbf{w} = \mathbf{\Phi}_L(\mathbf{\Phi}_L^T \mathbf{\Phi}_L + \zeta_1 \mathbf{B}_L^{-1})^{-1}\mathbf{D}_L \quad (36)$$

$$\begin{cases}
\boldsymbol{\Phi}_L = [\varphi_n,...,\varphi_{n+L-1}] = [\boldsymbol{\Phi}_{L-1}(2:L), \varphi_{n+L-1}] \\
\boldsymbol{\Theta}_L = [e(n),...,e(n+L-1)] = [\boldsymbol{\Theta}_{L-1}(2:L), e(n+L-1)]^T \\
\quad = [\boldsymbol{\Theta}_{L-1}(2:L), e_L]^T \\
[\boldsymbol{P}]_{ij} = \begin{cases} \left[\dfrac{\lambda\alpha_1}{L\beta_1^{\alpha_1}} G_{\alpha_1,\beta_1}(e(i))|e(i)|^{\alpha_1-2}\right], i=j \\ 0, i \neq j \end{cases} \\
[\boldsymbol{M}]_{ij} = \left[\dfrac{2(1-\lambda)\alpha_2}{L^2\beta_2^{\alpha_2}} G_{\alpha_2,\beta_2}(\Delta e_{i,j})|\Delta e_{i,j}|^{\alpha_2-2}\right] \\
\quad , i,j = n,...,n+L-1 \\
[\boldsymbol{N}]_{ij} = \begin{cases} \sum\limits_{k=n}^{n+L-1}\left[\dfrac{2(1-\lambda)\alpha_2}{L^2\beta_2^{\alpha_2}} G_{\alpha_2,\beta_2}(\Delta e_{i,k})|\Delta e_{i,k}|^{\alpha_2-2}\right], i=j \\ 0, i \neq j \end{cases} \\
\boldsymbol{B}_L = \boldsymbol{P}_L + \boldsymbol{M}_L - \boldsymbol{N}_L = \begin{bmatrix} \boldsymbol{B}_{L-1} & \boldsymbol{\Upsilon} \\ \boldsymbol{\Upsilon}^T & \psi_L \end{bmatrix} \\
[\boldsymbol{\Upsilon}]_i = \left[\dfrac{2(1-\lambda)\alpha_2}{L^2\beta_2^{\alpha_2}} G_{\alpha_2,\beta_2}(e_L - e(i))|e_L - e(i)|^{\alpha_2-2}\right], i=n,...,n+L-1 \\
\psi_L = \dfrac{\lambda\alpha_1}{L\beta_1^{\alpha_1}} G_{\alpha_1,\beta_1}(e_L)|e_L|^{\alpha_1-2} \\
\quad + \sum\limits_{k=n}^{n+L-2} \dfrac{2(1-\lambda)\alpha_2}{L^2\beta_2^{\alpha_2}} G_{\alpha_2,\beta_2}(e_L - e(k))|e_L - e(k)|^{\alpha_2-2}
\end{cases}$$
(37)

Therefore, from (36), convert the weight vector into a linear combination of $x_n$, i.e., $w = \boldsymbol{\Phi}_L \gamma_L$. Combining the first item of (36), $\boldsymbol{C}_L^{-1} = \boldsymbol{\Phi}_L^T \boldsymbol{\Phi}_L + \zeta_1 \boldsymbol{B}_L^{-1}$ is rewritten.

$$\boldsymbol{C}_L^{-1} = \begin{bmatrix} \boldsymbol{C}_{L-1}^{-1} & \boldsymbol{h}_L \\ \boldsymbol{h}_L^T & \varphi_L^T\varphi_L + \zeta_1\psi_L^{-1} \end{bmatrix} \quad (38)$$

(39) can be calculated with the block matrix inversion.

$$\boldsymbol{C}_L = r_L^{-1}\begin{bmatrix} \boldsymbol{C}_{L-1}r_L + \boldsymbol{z}_L\boldsymbol{z}_L^T & -\boldsymbol{z}_L \\ -\boldsymbol{z}_L^T & 1 \end{bmatrix} \quad (39)$$

where $\boldsymbol{z}_L = \boldsymbol{C}_{L-1}\boldsymbol{h}_L = \boldsymbol{C}_{L-1}\boldsymbol{\Phi}_{L-1}^T\varphi_L$ and $r_L = \varphi_L^T\varphi_L + \zeta_1\psi_L^{-1} - \boldsymbol{z}_L^T\boldsymbol{h}_L$.

Combing (39), (40) can be obtained

$$\begin{aligned}
\gamma_L &= \boldsymbol{C}_L \boldsymbol{D}_L \\
&= r_L^{-1}\begin{bmatrix} \boldsymbol{C}_{L-1}r_L + \boldsymbol{z}_L\boldsymbol{z}_L^T & -\boldsymbol{z}_L \\ -\boldsymbol{z}_L^T & 1 \end{bmatrix}\begin{bmatrix} \boldsymbol{D}_{L-1} \\ d_L \end{bmatrix} \\
&= \begin{bmatrix} \gamma_{L-1} - \boldsymbol{z}_L r_L^{-1} e_L \\ r_L^{-1} e_L \end{bmatrix}
\end{aligned} \quad (40)$$

The pseudocode of KGMEEF is shown in Table II.

TABLE II
THE ITERATIVE PROCESS OF THE KGMEEF ALGORITHM.

**Initialization:** $\boldsymbol{C}_1 = [\kappa(\boldsymbol{x}_1,\boldsymbol{x}_1) + \zeta_1(\dfrac{\lambda\alpha_1}{L\beta_1^{\alpha_1}} + \dfrac{2(1-\lambda)\alpha_2}{L\beta_2^{\alpha_2}})]^{-1}$,

$\gamma_1 = \boldsymbol{C}_1 d(1)$

**Parameters:** $\zeta_1, \lambda, \alpha_1, \beta_1, \alpha_2, \beta_2$

**Input:** data sequences $\{d(n), \boldsymbol{x}_n\}, n = 1, 2, ...$

**While** $\{d(n), \boldsymbol{x}_n\}$ available do

$\boldsymbol{h}_L = [\kappa(\boldsymbol{x}_{n+L-1}, \boldsymbol{x}_n), ..., \kappa(\boldsymbol{x}_{n+L-1}, \boldsymbol{x}_{n+L-2})]^T$

$\boldsymbol{z}_L = \boldsymbol{C}_{L-1}\boldsymbol{h}_L$

$r_L = \varphi_L^T\varphi_L + \zeta_1\psi_L^{-1} - \boldsymbol{z}_L^T\boldsymbol{h}_L$

$\boldsymbol{C}_L = r_L^{-1}\begin{bmatrix} \boldsymbol{C}_{L-1}r_L + \boldsymbol{z}_L\boldsymbol{z}_L^T & -\boldsymbol{z}_L \\ -\boldsymbol{z}_L^T & 1 \end{bmatrix}$

$y_L = \boldsymbol{h}_L^T \gamma_{L-1}$

$e_L = d_L - y_L$

$\gamma_L = \begin{bmatrix} \gamma_{L-1} - \boldsymbol{z}_L r_L^{-1} e_L \\ r_L^{-1} e_L \end{bmatrix}$

Compute $\psi_L$

**End**

### D. MLP Based on GMEEF

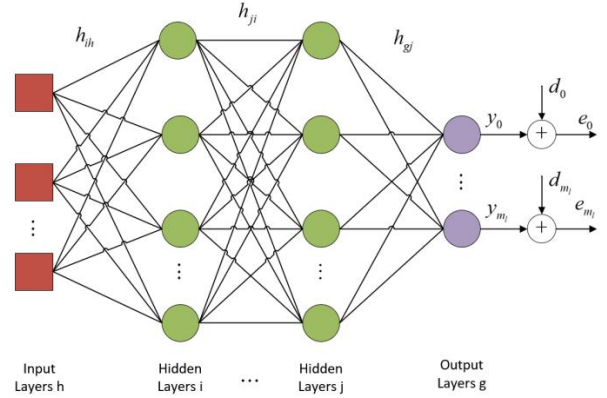

**Fig. 3.** A MLP Network.

MLP network follows a feedforward architecture. It comprises an input layer, one or more hidden layers, and an output layer. Within each layer, there are numerous neurons organized in a directed acyclic graph. The neurons in one layer are connected to one or more neurons in the subsequent layer, ensuring that information flows from the input layer to the output layer in a unidirectional manner [30].

MLPs are trained using supervised learning, where they learn to approximate a non-linear function that is useful for tasks like classification or regression. For instance, in the categorization problems, the input's features are transformed by the nonlinear activation function $\varphi(x)$ of the neurons in each layer. This transformation allows the data to be represented in a space where it can be linearly separated and classified into specific classes [31].

In this part, a novel MLP is presented by utilizing the new criterion (GMEEF) as the cost function of the weight iteration of each layer neurons. The cost function of GMEEF can effectively improve GMEE criterion's deficiencies.

The structure of the MLP is shown in Fig.3. In a MLP, each neuron receives inputs from one or more sources, and these inputs are assigned weights. Among these inputs, there is one called the bias, which always has a fixed value of one. The weighted inputs, along with the bias, are summed together to form a quantity known as the *net*. This *net* value is then passed through the $\varphi(x)$ (41) to produce an output $y$ (42).

$$\varphi(x) = \dfrac{1}{1+exp(-x)} \quad (41)$$

$$\begin{cases} net = \sum_{k=0}^{N} x_k h_k + b \\ y = \varphi(net) \end{cases} \quad (42)$$

The derivative of (41) is
$$\varphi'(x) = \varphi(x)(1-\varphi(x)) \quad (43)$$

The output $y$ from $k$ th neuron of $l$ th layer is represented as $y_k^l = \varphi(net_k^l)$. $m_l$ expresses the number of neurons in $l$ th layer. $h_{ki}^l$ denotes the connection weight between $i$ th neuron of $l-1$ th layer and $k$ th neuron of $l$ th layer.

Since the Parzen window estimation for the error PDF, the computation of the information potential involves considering all pairs of error samples. The total information potential, denoted as (44), is obtained by summing the individual information potentials of the output errors across all output neurons for a given sample pair $(\rho, \tau)$.

$$J(\rho,\tau) = \sum_{k=1}^{m_L} \left[ \frac{\lambda}{L} \sum_{\rho=1}^{L} G_{\alpha_1,\beta_1}(e_k(\rho)) + \frac{1-\lambda}{L^2} \sum_{\rho=1}^{L}\sum_{\tau=1}^{L} G_{\alpha_2,\beta_2}(e_k(\rho)-e_k(\tau)) \right]$$
$$= \sum_{k=1}^{m_L} [\lambda \widehat{V}_{\alpha_1,\beta_1}(e_k(\rho)) + (1-\lambda)\widehat{V}_{\alpha_2,\beta_2}(e_k(\rho)-e_k(\tau))] \quad (44)$$
$$= \sum_{k=1}^{m_L} [\widehat{V}_{GMEEF}(e_k(\rho)-e_k(\tau))]$$

The derivative of $J(\rho,\tau)$ with respect to $h_{gj}^l$ of the output layer using the gradient descent algorithm is as follows:

$$\frac{\partial J(\rho,\tau)}{\partial h_{gj}^l} = \lambda \frac{\partial \widehat{V}_{\alpha_1,\beta_1}(e_k(\rho))}{\partial e_k(\rho)} \frac{\partial e_k(\rho)}{\partial h_{gj}^l}$$
$$+ (1-\lambda) \frac{\partial \widehat{V}_{\alpha_2,\beta_2}(e_k(\rho)-e_k(\tau))}{\partial[e_k(\rho)-e_k(\tau)]} \left[ \frac{\partial e_k(\rho)}{\partial h_{gj}^l} - \frac{\partial e_k(s)}{\partial h_{gj}^l} \right]$$
$$= \lambda \frac{\partial \widehat{V}_{\alpha_1,\beta_1}(e_k(\rho))}{\partial e_k(\rho)} \frac{\partial e_k(\rho)}{\partial y_g^l(\rho)} \frac{\partial y_g^l(\rho)}{\partial net_g^l(\rho)} \frac{\partial net_g^l(\rho)}{\partial h_{gj}^l}$$
$$+ (1-\lambda) \frac{\partial \widehat{V}_{\alpha_2,\beta_2}(e_k(\rho)-e_k(\tau))}{\partial[e_k(\rho)-e_k(\tau)]} \begin{bmatrix} \frac{\partial e_k(\rho)}{\partial y_g^l(\rho)} \frac{\partial y_g^l(\rho)}{\partial net_g^l(\rho)} \frac{\partial net_g^l(\rho)}{\partial h_{gj}^l} \\ -\frac{\partial e_k(\tau)}{\partial y_g^l(\tau)} \frac{\partial y_g^l(\tau)}{\partial net_g^l(\tau)} \frac{\partial net_g^l(\tau)}{\partial h_{gj}^l} \end{bmatrix} \quad (45)$$
$$= -\lambda \varphi'(net_g^l(\rho)) y_j^{l-1}(\rho) \frac{\partial \widehat{V}_{\alpha_1,\beta_1}(e_k(\rho))}{\partial e_k(\rho)}$$
$$+ (1-\lambda) \frac{\partial \widehat{V}_{\alpha_2,\beta_2}(e_k(\rho)-e_k(\tau))}{\partial[e_k(\rho)-e_k(\tau)]} \begin{bmatrix} -\varphi'(net_g^l(\rho))y_j^{l-1}(\rho) \\ +\varphi'(net_g^l(\tau))y_j^{l-1}(\tau) \end{bmatrix}$$

where
$$\begin{cases} \frac{\partial \widehat{V}_{\alpha_1,\beta_1}(e_k(\rho))}{\partial e_k(\rho)} = \sum_{\rho=1}^{L} G_{\alpha_1,\beta_1}(e_k(\rho)) |e_k(\rho)|^{\alpha_1-1} sign(e_k(\rho)) \\ \frac{\partial \widehat{V}_{\alpha_2,\beta_2}(e_k(\rho)-e_k(\tau))}{\partial[e_k(\rho)-e_k(\tau)]} = \sum_{\rho=1}^{L}\sum_{\tau=1}^{L} \begin{bmatrix} G_{\alpha_2,\beta_2}(e_k(\rho)-e_k(\tau)) \times \\ |e_k(\rho)-e_k(\tau)|^{\alpha_2-1} sign(e_k(\rho)-e_k(\tau)) \end{bmatrix} \end{cases} \quad (46)$$

The derivative $J(\rho,\tau)/\partial h_{ji}^{l-1}$ of the hidden layer $l-1$ is as follows

$$\frac{\partial J(\rho,\tau)}{\partial h_{ji}^{l-1}} = \lambda \frac{\partial \widehat{V}_{\alpha_1,\beta_1}(e_k(\rho))}{\partial e_k(\rho)} \frac{\partial e_k(\rho)}{\partial h_{ji}^{l-1}}$$
$$+ (1-\lambda) \frac{\partial \widehat{V}_{\alpha_2,\beta_2}(e_k(\rho)-e_k(\tau))}{\partial[e_k(\rho)-e_k(\tau)]} \left[ \frac{\partial e_k(\rho)}{\partial h_{ji}^{l-1}} - \frac{\partial e_k(s)}{\partial h_{ji}^{l-1}} \right]$$
$$= \lambda \frac{\partial \widehat{V}_{\alpha_1,\beta_1}(e_k(\rho))}{\partial e_k(\rho)} \frac{\partial e_k(\rho)}{\partial y_j^{l-1}(\rho)} \frac{\partial y_j^{l-1}(\rho)}{\partial net_j^{l-1}(\rho)} \frac{\partial net_j^{l-1}(\rho)}{\partial h_{ji}^{l-1}}$$
$$+ (1-\lambda) \frac{\partial \widehat{V}_{\alpha_2,\beta_2}(e_k(\rho)-e_k(\tau))}{\partial[e_k(\rho)-e_k(\tau)]} \begin{bmatrix} \frac{\partial e_k(\rho)}{\partial y_j^{l-1}(\rho)} \frac{\partial y_j^{l-1}(\rho)}{\partial net_j^{l-1}(\rho)} \frac{\partial net_j^{l-1}(\rho)}{\partial h_{ji}^{l-1}} \\ -\frac{\partial e_k(\tau)}{\partial y_j^{l-1}(\tau)} \frac{\partial y_j^{l-1}(\tau)}{\partial net_j^{l-1}(\tau)} \frac{\partial net_j^{l-1}(\tau)}{\partial h_{ji}^{l-1}} \end{bmatrix} \quad (47)$$
$$= -\lambda \frac{\partial \widehat{V}_{\alpha_1,\beta_1}(e_k(\rho))}{\partial e_k(\rho)} \sum_{k=1}^{m_l} \begin{bmatrix} \varphi'(net_g^l(\rho))h_{kj}^l \\ \varphi'(net_j^{l-1}(\rho))y_i^{l-2}(\rho) \end{bmatrix}$$
$$+ (1-\lambda) \frac{\partial \widehat{V}_{\alpha_2,\beta_2}(e_k(\rho)-e_k(\tau))}{\partial[e_k(\rho)-e_k(\tau)]} \begin{bmatrix} -\sum_{k=1}^{m_l}\begin{bmatrix} \varphi'(net_g^l(\rho))h_{kj}^l \\ \varphi'(net_j^{l-1}(\rho))y_i^{l-2}(\rho) \end{bmatrix} \\ +\sum_{k=1}^{m_l}\begin{bmatrix} \varphi'(net_g^l(\tau))h_{kj}^l \\ \varphi'(net_j^{l-1}(\tau))y_i^{l-2}(\tau) \end{bmatrix} \end{bmatrix}$$

Like (46) and (47), the derivative of $J(\rho,\tau)$ on the bias $b^l$ of output layer and the bias $b^{l-1}$ of hidden layer $l-1$ are shown, respectively.

$$\begin{cases} \frac{\partial J(\rho,\tau)}{\partial b^l} = \lambda \frac{\partial \widehat{V}_{\alpha_1,\beta_1}(e_k(\rho))}{\partial e_k(\rho)} \varphi'(net_g^l(\rho)) \\ \quad + (1-\lambda) \frac{\partial \widehat{V}_{\alpha_2,\beta_2}(e_k(\rho)-e_k(\tau))}{\partial[e_k(\rho)-e_k(\tau)]} [\varphi'(net_g^l(\rho)) - \varphi'(net_g^l(\rho))] \\ \frac{\partial J(\rho,\tau)}{\partial b^{l-1}} = -\lambda \frac{\partial \widehat{V}_{\alpha_1,\beta_1}(e_k(\rho))}{\partial e_k(\rho)} \sum_{k=1}^{m_l} \begin{bmatrix} \varphi'(net_g^l(\rho))h_{kj}^l \\ \varphi'(net_j^{l-1}(\rho)) \end{bmatrix} \\ \quad + (1-\lambda) \frac{\partial \widehat{V}_{\alpha_2,\beta_2}(e_k(\rho)-e_k(\tau))}{\partial[e_k(\rho)-e_k(\tau)]} \begin{bmatrix} -\sum_{k=1}^{m_l}\begin{bmatrix}\varphi'(net_g^l(\rho))h_{kj}^l \\ \varphi'(net_j^{l-1}(\rho)) \end{bmatrix} \\ +\sum_{k=1}^{m_l}\begin{bmatrix}\varphi'(net_g^l(\tau))h_{kj}^l \\ \varphi'(net_j^{l-1}(\tau)) \end{bmatrix} \end{bmatrix} \end{cases} \quad (48)$$

IV. SIMULATION

In section IV, the computational complexity, and the application about GMEEF and QGMEEF criterion are calculated and tested.

*A. Computational Complexity*

TABLE III
COMPUTATIONAL COMPLEXITIES

| Algorithms | $\times/\div$ | $+/-$ | Exponentiation |
|---|---|---|---|
| LMS | $2M+1$ | $2M$ | 0 |
| LMF | $2M+1$ | $2M$ | 1 |
| GMCC | $2M+4$ | $2M+1$ | 3 |
| GMEE | $2M+ML+3L^2+3$ | $2M+ML+5L^2$ | $3L^2+2$ |
| GMEEF | $4M+ML+3L^2+7$ | $4M+ML+5L^2+1$ | $3L^2+5$ |
| QGMEEF | $4M+ML+4HL+8$ | $4M+ML+4HL+1+\sum_{i=1}^{L-1}i$ | $3HL+5$ |

Discuss the computational complexity of these new algorithms, which can effectively confirm the validity of the novel quantization algorithm and predict how fast the algorithm will run.

TABLE IV
$H_{ave}$ OF THE QGMEEF WITH DIFFERENT $\varepsilon$ ( $L = 100$ )

|  | $\varepsilon=0$ | $\varepsilon=0.01$ | $\varepsilon=0.03$ | $\varepsilon=0.05$ | $\varepsilon=0.1$ | $\varepsilon=0.3$ | $\varepsilon=0.5$ | $\varepsilon=0.8$ | $\varepsilon=1$ | $\varepsilon=5$ |
|---|---|---|---|---|---|---|---|---|---|---|
| Gaussian | 100 | 78.1 | 54.3 | 41.3 | 27.2 | 11.8 | 7.9 | 5.4 | 4.3 | 1.3 |
| Sub-Gaussian | 100 | 54.0 | 28.4 | 19.7 | 11.2 | 4.6 | 3.1 | 2.4 | 2.0 | 1.2 |
| Super-Gaussian | 100 | 34.6 | 20.6 | 14.3 | 11.5 | 8.0 | 6.0 | 5.8 | 5.2 | 3.4 |
| Rayleigh | 100 | 30.7 | 16.0 | 11.1 | 7.7 | 4.0 | 3.0 | 2.4 | 2.3 | 1.0 |

TABLE V
THE ITERATION REACH STEADY STATE, AND THE STEADY STATE VALUE (dBs) OF THE QGMEEF WITH DIFFERENT $\varepsilon$

|  | $\varepsilon=0$ | $\varepsilon=0.01$ | $\varepsilon=0.03$ | $\varepsilon=0.05$ | $\varepsilon=0.1$ | $\varepsilon=0.3$ | $\varepsilon=0.5$ | $\varepsilon=0.8$ | $\varepsilon=1$ | $\varepsilon=5$ |
|---|---|---|---|---|---|---|---|---|---|---|
| Gaussian | 259/-13.2 | 292/-13.8 | 411/-13.9 | 505/-14.2 | 634/-14.8 | 746/-15.8 | 886/-16.1 | 949/-16.5 | 1101/-16.7 | 1522/-17.6 |
| Sub-Gaussian | 291/-17.2 | 318/-16.9 | 342/-16.7 | 346/-16.7 | 384/-16.8 | 486/-16.6 | 502/-16.7 | 522/-16.8 | 540/-16.6 | 695/-16.7 |
| Super-Gaussian | 157/-25.7 | 181/-23.5 | 224/-21.6 | 235/-21.0 | 256/-20.3 | 316/-19.4 | 327/-19.7 | 341/-19.6 | 359/-19.6 | 384/-19.5 |
| Rayleigh | 137/-26.5 | 141/-23.8 | 137/-22.5 | 183/-21.6 | 205/-21.2 | 218/-20.5 | 220/-20.6 | 224/-20.4 | 233/-20.5 | 247/-20.6 |

The computational complexity of these new algorithms, GMEEF and QGMEEF, is compared with several existing LMS, LMF, GMCC, and GMEE algorithms, as shown in Table III. To value the total computation of these algorithms, the method proposed by Chen et al. [32] is employed, which involves summing up the computations for each algorithm. The results are displayed as follows

Table III reveals that the GMEEF algorithm exhibits a higher computational complexity when compared to other algorithms. Nevertheless, it is worth mentioning that the quantization mechanism employed in the GMEEF algorithm utilizes a considerably smaller number of real-valued codes $H$ compared to the length of the sliding window $L$. This implies that the integration of the quantization mechanism actively alleviates the computational burden of the GMEEF algorithm, resulting in minimal adverse effects on its overall performance.

In summary, the quantization mechanism employed in the QGMEEF algorithm helps to lower the computational complexity compared to the original GMEEF algorithm, while maintaining comparable performance. This indicates that the algorithm is expected to run faster without sacrificing its effectiveness.

*B. Simulation of Linear System Identification*

In this part, Fig.4, Table IV and Table V show the effect of parameters $L$ and $\varepsilon$ change on the performance of GMEEF and QGMEEF, where $H_{ave}$ is the average of $H$ in online VQ method. It has a reference value for the selection of $L$ and $\varepsilon$ in the subsequent test performance. Since QGMEEF type algorithm has a denser distribution of MSD curves with different $\varepsilon$, the variation of MSD curves will be represented by Tables IV and V. In Fig.4, as $L$ increases, the convergence rate of these new algorithms is almost inconvenient, but the steady-state error then decreases and the computational complexity increases accordingly. In Table IV, as $\varepsilon$ increases, $H_{ave}$ decreases accordingly, the QGMEE part in the QGMEEF loses some parameters and the performance decays accordingly, while the influence of GMCC part in the QGMEEF is enhanced accordingly, and the MSD curve shows a trend of slowing down the convergence rate and decreasing the steady-state value. Therefore, the choice of parameters is particularly important.

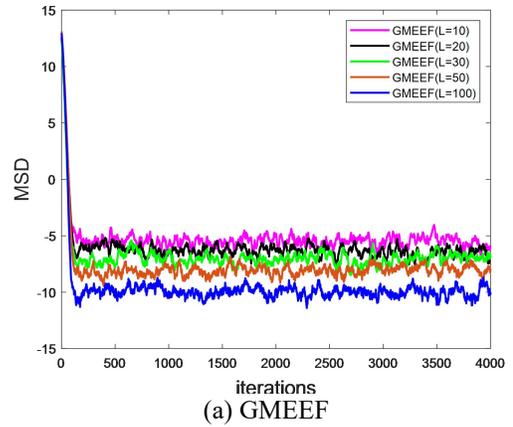

(a) GMEEF

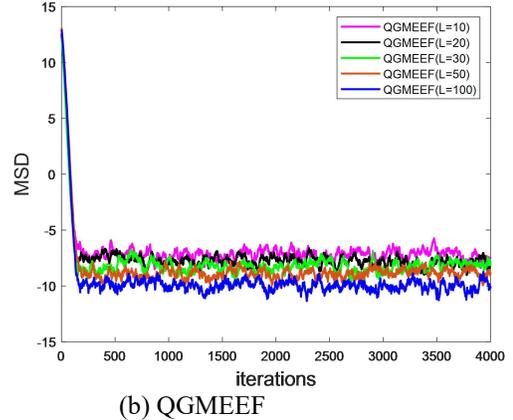

(b) QGMEEF

**Fig. 4.** The comparison of the algorithms' performance with different parameter $L$ under Gaussian noise, where $\mu = 0.1, \alpha_1 = 2, \beta_1 = 10, \alpha_2 = 1, \beta_2 = 20, \lambda = 0.8, \varepsilon = 0.02$.

Next, to showcase the exceptional performance of the GMEEF and QGMEEF algorithms, four different noise distributions are employed. The evaluation considers both the computational workload and the steady-state performance of these new algorithms, $L = 50$ and $\varepsilon = 0.02$ are utilized.
1) Consider the existence of a Gaussian noise with zero-mean and unit variance.
2) Consider the existence of a Sub-Gaussian noise follows

[33] distribution.
3) Consider the existence of a Super-Gaussian noise follows mixed Gaussian distribution with $v(n) \sim 0.95N(0,0.01) + 0.05N(0,100)$.
4) Consider the existence of a Super-Gaussian noise follows zero-mean Rayleigh distribution. The PDF [34] is denoted as $v(t) = (t/\sigma^2)\exp(-t^2/2\sigma^2)$, where $\sigma = 3$.

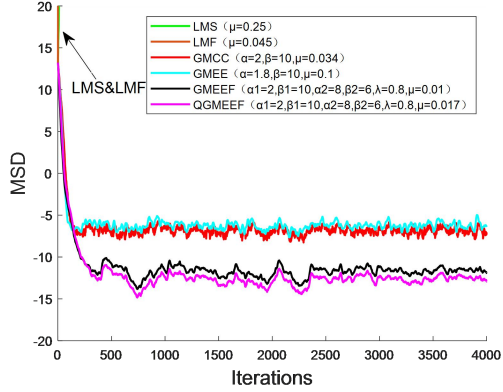

(a) Gaussian noise

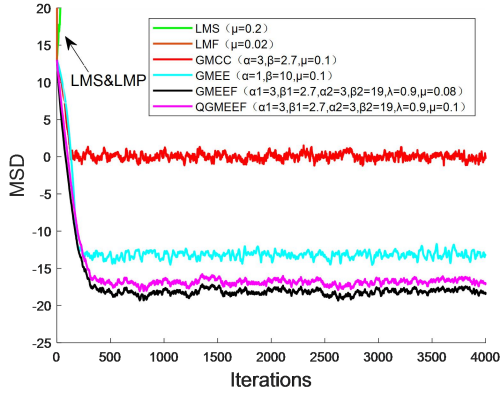

(b) Sub-Gaussian noise

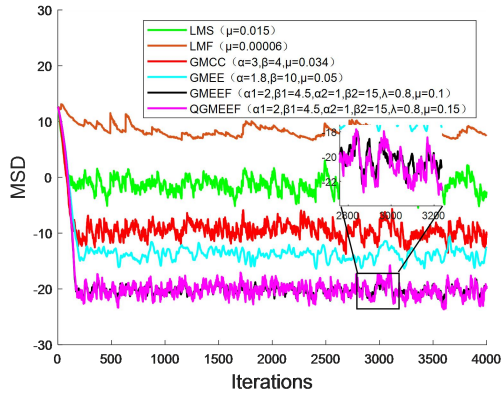

(c) Mixed Gaussian noise

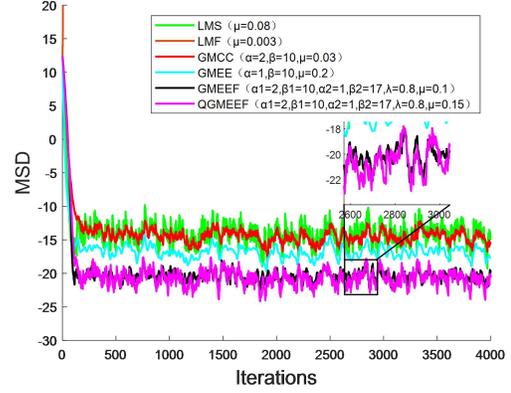

(d) Rayleigh noise

**Fig. 5.** The comparison of the algorithms' performance under different noises.

In Fig.5, the proposed GMEEF and QGMEEF always keep the most excellent performance than the other algorithms. Theoretically, the quantized algorithm should approximate or slightly outperform the unquantized algorithm in terms of steady-state performance and convergence rate, which is fully demonstrated in Fig. 5 (a), (c), (d). However, observing Fig. 5 (b), it can see that the performance of the QGMEEF is a little worse compared to GMEEF, which may be the reason that the GMEEF only quantizes the GMEE part of it. And the GMEE part has a bad performance in the sub-Gaussian noise.

*C. The Simulation of Acoustic Echo Canceller*

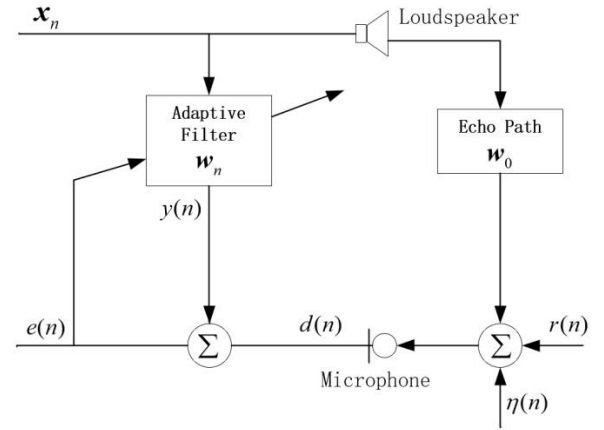

**Fig. 6**. The structure of AEC.

Fig. 6 illustrates the structure of AEC system. AEC employs the pseudo echo generated by the AF to estimate the impulsive response of the acoustic echo path. This estimated response is subtracted from the actual echo to mitigate its presence. During this process, the AF temporarily halts its coefficient updating. However, in the presence of double-talk (simultaneous speech from both ends), the quantity of the speech signal can become more pronounced or exacerbated. Thus, in the presence of double-talk (simultaneous speech from both ends), the accuracy of echo path estimation may be compromised [35]. To ensure correct convergence of the AF, a double-talk detector is employed.

In Fig. 6, $e(n)$ denotes the error signal. $y(n)$ indicates the estimated echo signal by AF. $v(n)$ expresses the near end talk. $\eta(n)$ represents the background noise.

For evaluating the performance of the proposed algorithms, the ICASSP dataset [36] is used for training purposes. This dataset is created by synthesizing speech recordings, room impulsive responses, and background noise obtained from [37]. Fig. 7 illustrates the far-end and near-end speech signals present in the dataset.

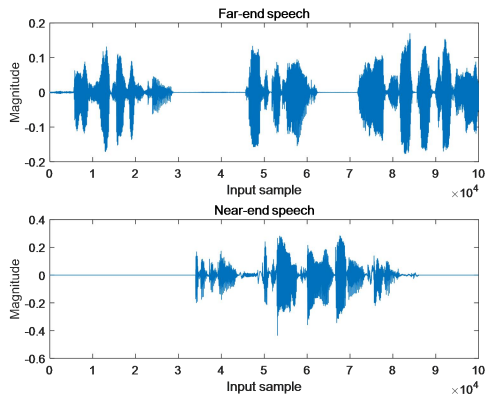

**Fig. 7.** Speech Signals.

For the far-end signal, a random speaker is selected from a pool of 1,627 speakers, and a 10-second audio clip is sampled from their recordings. For the near-end signal, another speaker is randomly chosen, and 3-7 seconds of audio are zero-padded to match the length of the far-end signal. For generating the echo, a random room impulsive response is chosen from a database, with a range of RT60 values from 200 ms to 1200 ms, using Project Acoustics technology. In 80% of the cases, the far-end signal undergoes nonlinear distortion to mimic loudspeaker characteristics. The echo is mixed with the near-end signal at various signal-to-echo ratios ranging from -10 dB to 10 dB. The echo return loss enhancement (ERLE), $10\log_{10} E[d^2(n)]/E[e^2(n)]$, dB, as a tool, measure the algorithm performance. $E[d^2(n)]$ and $E[e^2(n)]$ is recursively estimated by (49) with $\chi = 0.999$ [38].

$$\begin{cases} \varpi_d^2(n) = \chi \varpi_d^2(n-1) + (1-\chi)d^2(n) \\ \varpi_e^2(n) = \chi \varpi_e^2(n-1) + (1-\chi)e^2(n) \end{cases} \quad (49)$$

From the dataset, random samples of far-end and near-end speech are selected. The performance of the proposed algorithms is evaluated and depicted in Fig. 8, demonstrating their excellent performance.

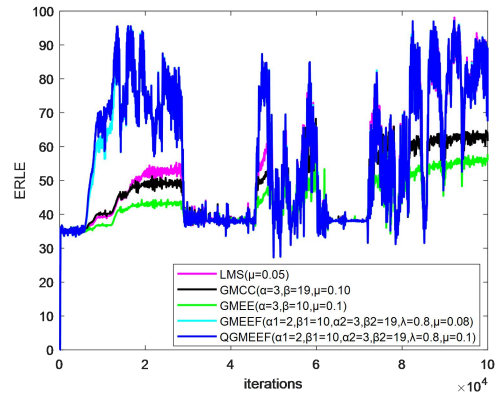

**Fig.8.** The different ERLE curves with a double-talk scenario.

*D. The Simulation of Time Series Prediction*

In part D, Mackey – Glass (MG) chaotic time series are utilized to evaluate the nonlinear learning capability of the KRGMEEF algorithm. The MG equation can be expressed in the form of [24]

$$\frac{dx(t)}{dt} = \frac{0.2x(t-\Delta)}{1+x^{10}(t-\Delta)} - 0.1x(t) \quad (50)$$

The 1000 training data with a noise (this noise satisfies the (3) and (4) distribution of part B of Chapter IV, respectively). The performance of KRGMEEF is compared with KRGMCC [29], KRGMEE [26] is demonstrated by convergence curves of mean square error (MSE) in Fig.9 (a) and (b). In Fig.9, it is obvious that the performance of KRGMEEF under mixed Gaussian noise and Rayleigh noise is excellent.

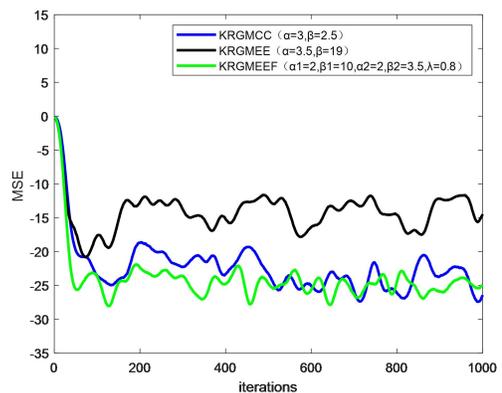

(a) Mixed gaussian noise

TABLE V
PARAMETERS, TRAINING AND TESTING ACCURACIES FOR DIFFERENT COST FUNCTIONS UNDER MNIST DATASET.

| Algorithms | $\alpha_1$ | $\beta_1$ | $\alpha_2$ | $\beta_2$ | $\lambda$ | Training Accuracy | Testing Accuracy |
|---|---|---|---|---|---|---|---|
| MLP-CE | - | - | - | - | - | 0.9591 | 0.9443 |
| MLP-GMCC | 2 | 1.5 | - | - | - | 0.9598 | 0.9517 |
| MLP-GMEE | - | - | 3.5 | 6 | - | 0.9614 | 0.9082 |
| MLP-GMEEF | 2 | 1.5 | 2.5 | 3 | 0.8 | 0.9705 | 0.9601 |

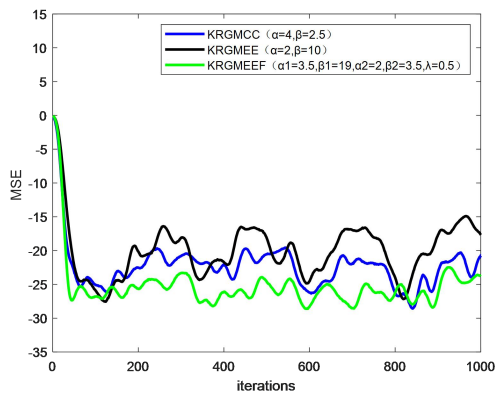

(b) Rayleigh noise

**Fig.9.** The MSE comparative curves in data forecasting.

*E. The Simulation of Supervised Classification*

The performance of supervised classification is utilized the MLP network to test. The dataset utilizes the MNIST handwriting dataset [39]. The MNIST dataset contains 784 features, 10000 training samples, and 1000 testing samples. A 4-layers MLP is used to train and classify these handwriting numbers. The input nodes, and output nodes are corresponding to input feature and classification species, respectively. The hidden layer 1 nodes and hidden layer 2 nodes are 300 and 100, respectively. The MLP algorithms based on cross entropy (MLP-CE), GMCC, GMEE, GMEEF are compared, and the parameters, training accuracies, and testing accuracies are shown in Table V. In the simulation, the MLP-GMEEF also shows excellent performance.

## VI. CONCLUSION

In conclusion, the GMEEF criterion is proposed to solve the drawback of the fixed gaussian kernel of MEEF. The properties of GMEEF are simply discussed and QGMEEF is presented to lower high computational complexity of GMEEF. They stabilization steps range are calculated. Secondly, from the performance comparison of these criterions in system identification and AEC, it is obvious that the GMEEF and QGMEEF have the similar performance, and show the more excellent performance than the other algorithms. Moreover, to handle the nonlinear signals, the KRGMEEF algorithm is presented. And the excellent prediction of the KRGMEEF algorithm and its superior performance are verified by forecasting the MG time series. Finally, the GMEEF criterion is applied to the MLP network and tested the promising application of the criterion to supervised classification. The GMEEF-type algorithm has more excellent performance than other algorithm, but relatively, there are also many more variable parameters, so how to reduce the variable parameters in the future is a worthy research direction.